# A Fractional Differential Transformation Solution Method for the Assessment, Monitoring, Control, and Evaluation of HIV/AIDS Confirmed Status with Vertical Transmission in Nigeria

**Capt. (NN) M. O. Oladejo**[1] **Ali Onuche John**[2]
Department of Mathematical Sciences, Nigerian Defence Academy, Kaduna
1. Mikeoladejo2003@yahoo.com 2. Alionuche26@gmail.com

## ABSTRACT
HIV is a deadly virus transmitted either through having of unprotected sex, mother to child transmission, sharing of unsterilized objects that is capable of making cut or wounds on the body, through blood or bodily fluid transmission. AIDS has no permanent cure but remedies that help in suppressing the power effect of the virus are available. Previous studies has shown that the epidemic claimed more lives via vertical transmission, mother-to-child transmission, blood transfusion, sexual intercourse and injection drug users. The associated models were derived using diagnosis and treatments records of confirmed status of HIV/AIDS clients. A new model analysis were proposed that could handle cases of HIV/AIDS routes of transmission and treatments via vertical, mother-to-child, blood transfusion, sexual intercourse, injection drug users that were not considered in previous studies. The new model was solved using Fractional Differential Transformation by Caputo sense algorithm. The new model is a major contribution for optimal assessment, monitoring, evaluation, control and management of HIV/AIDS, with 75% accuracy.

## 1. INTRODUCTION

"HIV and AIDS are global catastrophe, the biggest plague, and the worst Tsunami in human history [14]". The Joint United Nations Programme on HIV/AIDS [13] has nicknamed HIV/AIDS as the most devastating disease that mankind has ever face. Similarly in the global geopolitical world HIV/AIDS has been categorized under the five major causes of death in the world. The pandemic of HIV epidemic right from its reception to date has been a global blow, challenge, an intriguing burden and has presented a lot of serious concern and problems to public health workers, most especially in developing countries especially in Nigeria. Categorically HIV is of two types; HIV-1 and HIV-2. The former is the most widespread worldwide while the latter is a less prevalent and less pathogenic and is majorly found principally in West African countries. A model that handles some of the cases through which HIV/AIDS is been transmitted was developed using Nigeria as a case study.

The Federal Republic of Nigeria commonly referred to as Nigeria is a Federal Constitutional Republic in West Africa bordering Benin, in the west, Chad and Cameron in the east, and Niger in the north. Its coast in the south lies on the Gulf of Guinea in the Atlantic Ocean. It comprises of 36 states and the Federal Capital Territory (FCT) Abuja located in the middle-belt of Nigeria with Lagos been the largest populous and most industrious city in the country. The three major languages speak in Nigeria are Hausa, Yoruba and Igbo. The major occupation in Nigeria is farming. Worldwide Nigeria is among the top ten most populous countries in the world and the highest populated country in Africa. Naturally bless with oil and other natural resources. It has a population of about 187, 179, 898 population [18]. But sadly enough HIV/AIDS has been a major challenge in this country, 210,000 people die annually in this country because of HIV/AIDS and to be so pathetic enough 3.2 million people are living with HIV with adult prevalence of 3.2%, 400,000 children are born with HIV and a total sum of 220,000 new infections has been recorded so far from Nigeria alone, thus making Nigeria the second country with the highest cases of HIV/AIDS. But painful and more sorrowful enough only 21% of the total sum of people living with HIV/AIDS has access to receives treatments [4], this is due to either;

i) Poverty: most Nigerians are poor, they don't even have the money to provide three-square meal for themselves talk more of having the means of affording money to purchase those costly anti-retroviral drugs (ART).
ii) Ignorance: Nigerians don't believe in taking medications when sick, they prefer managing the condition or taking herbal medicine
iii) Shy: Nigerians are very good in stigmatizing and criticizing someone, so in fear of this people living with HIV hides their status identity and pretends everything is normal with them.
iv) Negligence: To the average man of Nigeria sickness is a show of laziness, thus when they are sick that's when they tends to ignorantly proves themselves to be man by not taking drugs or treatments until the situation overpower them before they start thinking of drugs, treatments or solution.

All this aforementioned above contributes to the major reasons while Nigeria has the highest number of AIDS-related death cases in the world.

Routes through which HIV/AIDS spread in Nigeria has been boiled down to; unprotected sexual intercourse, blood transfusion without proper examination, mother-to-child transmission either during pregnancy, delivery or breast feeding a baby, exchange of needles among injection drug users, careless making use of unsterilized sharp objects especially razor blades and clippers,

People who are in great danger of contacting the virus include;

i) Men who have sex with men (Homosexual men or Gay).
ii) Women who have sex with women (Homosexual women or Lesbians).
iii) Ladies who have sex with men in exchange of money (Ashawo, Olosho, Runs Girls or Prostitutes).
iv) Young guys who are engaged in sex with a sugar mummy (Gigolo).
v) Married men who cheat on their wife by sleeping around with young ladies.
vi) Women who cheats on their husband by sleeping around with their boss or rich men.
vii) Pregnant women especially if the status of their partner is unknown.
viii) Children conceived by an infected mother.
ix) People who take hard drugs.
x) Victims of rape.
xi) Prison inmates
xii) Adventure in sex (youths explore, adults pleasure, and show-off etc.)

Though awareness and enlightenment campaign have been mounted, but these have not significantly curbed the danger. In order to drastically and effectively control the menace a mathematical model is been formulated for optimal control and management.

## 2. LITERATURE

Mathematical modeling involves using of symbols, semantics, flowcharts, and algebraic functions and relations etc. to represents a system of abstraction that tends to imitate the dynamics or represents nature or situation of a system at hand. Many scholars Anderson and May [11], Mary et. al., [10], Chomosky et. al., [5], Naresh et. al., [12], Isack et. al., [8] and Abdalla et. al., [1] has used mathematical model in modeling the transmission of HIV/AIDS in a time-varying population size. All their models are case and specific to one or two major routes of HIV/AIDS transmission mechanism. Ways through which the virus is been transmitted is listed in [3]. A model that handles such transmission is been developed in this paper and using Fractional Differential Transformation in Caputo sense the model was solved.
A function is said to be in Caputo sense if

$$D^{\alpha}x^{n} = \left\{\begin{array}{ll} 0 ----------------- for\ n\epsilon N_0\ and\ n < \lceil\alpha\rceil \\ \dfrac{\Gamma(n+1)}{\Gamma(n+1-\alpha)}x^{n-\alpha} ---------- for\ n\epsilon N_0\ and\ n \geq \lceil\alpha\rceil \end{array}\right\} ----- 2.13$$

Where $\lceil\alpha\rceil$ is a ceiling function which denotes the smallest integer than or equal to $\alpha$. $N_0 = \{0,1,2,3, \ldots\ldots\}$, $\Gamma$ denote gamma function, $D^{\alpha}x^{n}$ represents Caputo fractional derivative of order n for more details, Ahmed et al.,[2]. Ahmed used Caputo sense to solve the SIR model, S.Z Rida [6] developed an epidemic model system and analysis of a non-fatal diseases using the Caputo sense, Hanaa and Mohammed [7] applied Caputo sense in solving an MSEIR epidemic model. In this paper we are using the Caputo sense to solve our model of assessments monitoring, control, and evaluation of HIV/AIDS confirmed status with vertical transmission in Nigeria.

## 3. METHODOLOGY

We divide the population size of Nigeria into five different compartments. Those without the virus susceptible S(t), those having HIV virus I(t), those having AIDS A(t), those in Pre-AIDS P(t) and those receiving treatments T(t). Other parameters of the model are presented in the table below.

Table 1: Parameter data.

| | |
|---|---|
| $\beta$ | Probability of HIV/AIDS transmission<br>Where the subscripts<br>xx: rate of transmission from man to man<br>mm: rate of transmission from man to man<br>xy: rate of transmission from man to woman.<br>yx: rate of transmission from woman to man<br>xx: rate of transmission from woman who to woman<br>ax: anal sex (man-man)<br>xa: Anal sex man-woman<br>b: blood transfusion<br>m: Injection stick injury |
| | f: fellatio receptive<br>d:injection drug users |
| $N$ | Total population size. Where the subscripts m and f represents the total population size for male and female respectively. |
| $S$ | Susceptible individuals |
| $I$ | Infected individuals |
| $T$ | Individuals of antiretroviral (ART) drugs |
| $A$ | AIDS individuals |
| $P$ | Individuals who are positive but has not started taking ART. |
| $\mu$ | Death rate |
| $\pi$ | Birth rate |
| $\gamma$ | Rate of movement of clients on care treatments. |
| $\varepsilon$ | Probability of movement from I |
| $\sigma$ | Rate of movement from I |
| $C$ | Rate of movement from S. |
| $k$ | Rate of movement from A to T |
| $v$ | Rate of movement from T to A. |
| $d$ | Death due to AIDS virus. |
| $\theta$ | Probability of HIV/AIDS transmission from mother to child |
| $\tau$ | Rate of transmission of HIV/AIDS from mother to child |

The information given in the table above is used to generate the transition network diagram below.

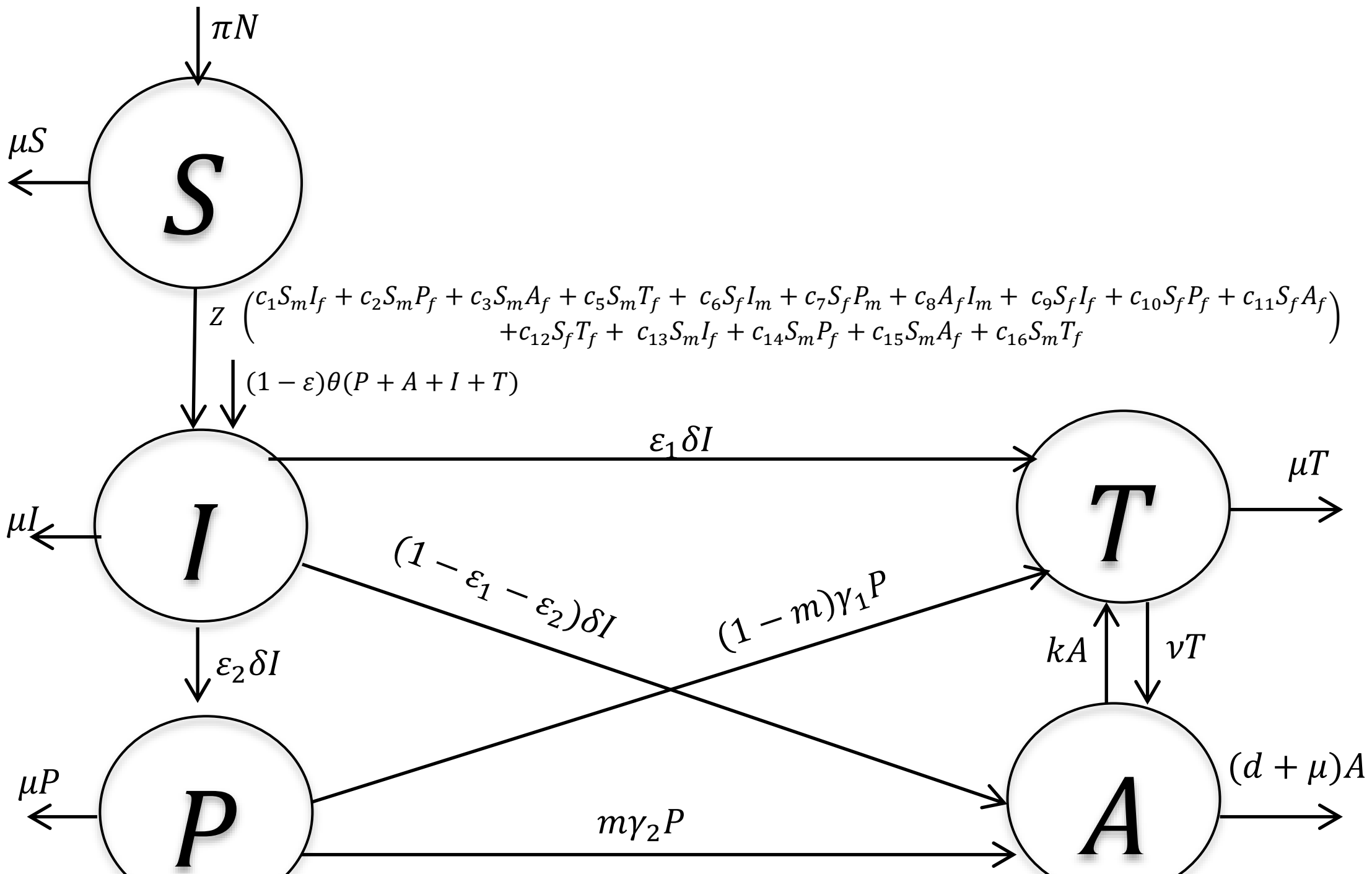

Figure 1: Flow chart for the proposed model

Figure.1 is the network diagram of the proposed model

Where

$$Z = \frac{(\beta_{xx} + \beta_{xy} + \beta_{yx} + \beta_{yy} + \beta_{ax} + \beta_{xa} + \beta_{ay} + \beta_d + \beta_b + \beta_{jm} + \beta_f)}{N}$$

## Explicit Proposed Generalized Model

$$\frac{dS}{dt} = \pi N - \frac{(\beta_{xx} + \beta_{xy} + \beta_{yx} + \beta_{yy} + \beta_{ax} + \beta_{xa} + \beta_{ay} + \beta_d + \beta_b + \beta_{jm} + \beta_f)}{N}\begin{bmatrix} c_1 S_m I_f + c_2 S_m P_f + c_3 S_m A_f + c_5 S_m T_f \\ + c_6 S_f I_m + c_7 S_f P_m + c_8 A_f I_m \\ + c_9 S_f I_f + c_{10} S_f P_f + c_{11} S_f A_f + c_{12} S_f T_f \\ + c_{13} S_m I_f + c_{14} S_m P_f + c_{15} S_m A_f + c_{16} S_m T_f \end{bmatrix} - \mu S - - - - 1a$$

$$\frac{dI}{dt} = \frac{(\beta_{xx} + \beta_{xy} + \beta_{yx} + \beta_{yy} + \beta_{ax} + \beta_{xa} + \beta_{ay} + \beta_d + \beta_b + \beta_{jm} + \beta_f)}{N}\begin{bmatrix} c_1 S_m I_f + c_2 S_m P_f + c_3 S_m A_f + c_5 S_m T_f \\ + c_6 S_f I_m + c_7 S_f P_m + c_8 A_f I_m \\ + c_9 S_f I_f + c_{10} S_f P_f + c_{11} S_f A_f + c_{12} S_f T_f \\ + c_{13} S_m I_f + c_{14} S_m P_f + c_{15} S_m A_f + c_{16} S_m T_f \end{bmatrix}$$

$$+(1-\theta)\tau(I + P + A + T) - \varepsilon_2 \delta I - \varepsilon_1 \delta I - (1 - \varepsilon_1 - \varepsilon_2)\delta I - \mu I \quad - - - - - - - - - - - - - - - - 1b$$

$$\frac{dP}{dt} = \varepsilon_2 \delta I - (1-m)\gamma_2 P - m\gamma_1 P - \mu P - - - - - - - - - - - - - - - - - - - - - - - - - 1c$$

$$\frac{dA}{dt} = (1 - \varepsilon_1 - \varepsilon_2)\delta I + vT + (1-m)\gamma_2 P - kA - (\mu + d)A - - - - - - - - - - - - - - 1d$$

$$\frac{dT}{dt} = \varepsilon_1 \delta I + m\gamma_1 P + kA - vT - \mu T - - - - - - - - - - - - - - - - - - - - - - - - - - 1e$$

$$- - - (1)$$

For simplicity of the model, let all the infective be represented by I, all the susceptible by S, al the Pre-AIDS by P, all those receiving treatments be T and those with AIDS be A. equation 1 will become

$$\frac{dS}{dt} = \pi N - \frac{(\beta_{xx} + \beta_{xy} + \beta_{yx} + \beta_{yy} + \beta_{ax} + \beta_{xa} + \beta_{ay} + \beta_d + \beta_b + \beta_{jm} + \beta_f)}{N}[C_1 SI + C_2 SP + C_3 ST + C_4 SA] - \mu S - - - - - 2a$$

$$\frac{dI}{dt} = \frac{(\beta_{xx} + \beta_{xy} + \beta_{yx} + \beta_{yy} + \beta_{ax} + \beta_{xa} + \beta_{ay} + \beta_d + \beta_b + \beta_{jm} + \beta_f)}{N}[C_1 SI + C_2 SP + C_3 ST + C_4 SA]$$

$$+(1-\theta)\tau(I + P + A + T) - \varepsilon_2 \delta I - \varepsilon_1 \delta I - (1 - \varepsilon_1 - \varepsilon_2)\delta I - \mu I \quad - - - - - - - - - - - - - - - - 2b$$

$$\frac{dP}{dt} = \varepsilon_2 \delta I - (1-m)\gamma_2 P - m\gamma_1 P - \mu P - - - - - - - - - - - - - - - - - - - - - - - - - 2c$$

$$\frac{dA}{dt} = (1 - \varepsilon_1 - \varepsilon_2)\delta I + vT + (1-m)\gamma_2 P - kA - (\mu + d)A - - - - - - - - - - - - - - 2d$$

$$\frac{dT}{dt} = \varepsilon_1 \delta I + m\gamma_1 P + kA - vT - \mu T - - - - - - - - - - - - - - - - - - - - - - - - - - 2e$$

$$(2)$$

$Theorem\ 1: let\ \Omega = \{(s, i, p, h, a) \epsilon\ \mathbb{R}_+^5 : s + i + p + a + h = 1\},$
$then\ the\ solutions\ \{s(t), i(t), p(t), a(t), h(t)\}\ of\ the\ system\ are\ positive\ \forall t \geq 0.$

Prove:

$$\frac{ds}{dt} \geq \pi - s\pi$$

Integrating

$$s = 1 + ce^{-\pi t}$$
$$as\ t \to 0$$
$$c = s_0 - 1$$
$$as\ t \to \infty, \quad s = 1$$
$$\Longrightarrow 0 \leq s(t)$$

$$\frac{di}{dt} \leq \theta i(1 - \varepsilon) - (1 - \varepsilon_2)\delta + \pi$$

By integrating and checking for $t \to 0\ \ and\ \ t \to \infty \quad i(t) \geq 0$

$$\frac{dp}{dt} \le -(1-m)\gamma_2 + \pi + m\gamma_1)p$$

By integrating and checking for $t \to 0$ $and$ $t \to \infty$ $p(t) \ge 0$

$$\frac{dh}{dt} \le -(v+\pi)h$$

By integrating and checking for $t \to 0$ $and$ $t \to \infty$ $h(t) \ge 0$

$$\frac{da}{dt} \le -(k+\pi)a$$

By integrating and checking for $t \to 0$ $and$ $t \to \infty$ $a(t) \ge 0$ $QED.$

In contrary to Abdalla's assumption, we assume that both Pre-AIDS and AIDS patients contribute to the spread of the virus both horizontally and vertically, thus, the total population size of the system is

$$\left.\begin{aligned} &N = S+E+I+P+A+T \text{ -------------- } 3a \\ &\frac{dN}{dt} = \frac{dS}{dt}+\frac{dE}{dt}+\frac{dI}{dt}+\frac{dP}{dt}+\frac{dA}{dt}+\frac{dT}{dt} \text{ --------- } 3b \\ &\frac{dN}{dt} = \pi N + (1-\varepsilon)\theta(I+P+A+T) - \mu N - dA \text{ -------- } 3c \end{aligned}\right\} \text{ ------ } (3)$$

Taking the system average for each compartment

$$\left.\begin{aligned} s &= \frac{S}{N} \text{ --------- } (4a) \\ i &= \frac{I}{N} \text{ --------- } (4b) \\ e &= \frac{E}{N} \text{ --------- } (4c) \\ p &= \frac{P}{N} \text{ --------- } (4d) \\ a &= \frac{A}{N} \text{ --------- } (4e) \\ h &= \frac{T}{N} \text{ --------- } (4f) \end{aligned}\right\} \text{ --------- } (4)$$

From equation (4)

$$\left.\begin{aligned} &S = sN \text{ ------------ } 5a \\ &\textit{Applying Product rule} \\ &\frac{dS}{dt} = s\frac{dN}{dt} + N\frac{ds}{dt} \text{ ---------- } 5b \\ &\frac{ds}{dt} = \frac{1}{N}\left[\frac{dS}{dt} - s\frac{dN}{dt}\right] \text{ ---------- } 5c \\ &\textit{in similar vein} \\ &\frac{di}{dt} = \frac{1}{N}\left[\frac{dI}{dt} - i\frac{dN}{dt}\right] \text{ ---------- } 5d \\ &\frac{de}{dt} = \frac{1}{N}\left[\frac{dE}{dt} - e\frac{dN}{dt}\right] \text{ ---------- } 5e \\ &\frac{dp}{dt} = \frac{1}{N}\left[\frac{dP}{dt} - p\frac{dN}{dt}\right] \text{ ---------- } 5f \\ &\frac{da}{dt} = \frac{1}{N}\left[\frac{dA}{dt} - a\frac{dN}{dt}\right] \text{ ---------- } 5g \\ &\frac{dh}{dt} = \frac{1}{N}\left[\frac{dT}{dt} - h\frac{dN}{dt}\right] \text{ ---------- } 5h \end{aligned}\right\} \text{ ------- } (5)$$

Substituting (1) into (4)

$$\frac{ds}{dt} = \pi - \frac{(\beta_{xx} + \beta_{xy} + \beta_{yx} + \beta_{yy} + \beta_{ax} + \beta_{xa} + \beta_{ay} + \beta_d + \beta_b + \beta_{jm} + \beta_f)}{N}[C_1 si + C_2 sp + C_3 sh + C_4 sa] - \mu s$$
$$-s(\pi + (1-\varepsilon)\theta(i+p+a+h) - \mu - da) - - - - - 6a$$

$$\frac{di}{dt} = \frac{(\beta_{xx} + \beta_{xy} + \beta_{yx} + \beta_{yy} + \beta_{ax} + \beta_{xa} + \beta_{ay} + \beta_d + \beta_b + \beta_{jm} + \beta_f)}{N}[C_1 si + C_2 sp + C_3 sh + C_4 sa]$$

$$+(1-\varepsilon)\theta(i+p+a+h) - \varepsilon_2 \delta i - \varepsilon_1 \delta i - (1-\varepsilon_1-\varepsilon_2)\delta i - \mu i - i(\pi + (1-\varepsilon)\theta(i+p+a+h) - \mu - da) \quad - -65b$$

$$\frac{dp}{dt} = \varepsilon_2 \delta i - (1-m)\gamma_2 p - m\gamma_1 p - \mu p - p(\pi + (1-\varepsilon)\theta(i+p+a+h) - \mu - da) - - - - - - - - - - 6c$$

$$\frac{da}{dt} = (1-\varepsilon_1-\varepsilon_2)\delta i + vh + (1-m)\gamma_2 p - ka - (\mu + d)a - a(\pi + (1-\varepsilon)\theta(i+p+a+h) - \mu - da) - - - - 6d$$

$$\frac{dh}{dt} = \varepsilon_1 \delta i + m\gamma_1 p + ka - vh - \mu h - h(\pi + (1-\varepsilon)\theta(i+p+a+h) - \mu - da) - - - - - - - - - - - - - - - - - 6e$$

$$(6)$$

The equilibrium point of system 5 can be obtained by setting

$$\frac{ds}{dt} = \frac{di}{dt} = \frac{dp}{dt} = \frac{dh}{dt} = \frac{da}{dt} = 0 - - - - - - - - 6f$$

The disease-free-equilibrium is obtained when

$$i = p = h = a = 0.$$

Thus

$$0 = \pi - s^*(\pi + \mu) - - - - - - - - 7$$

Producing

$$s^* = \frac{\pi}{\pi + \mu} - - - - - - - - 8$$

Thus the disease-free-equilibrium

$$E^0 = \left\{\frac{\pi}{\pi + \mu}, 0,0,0,0\right\} - - - - - - 9$$

The endemic disease equilibrium

$$E^* = \{s^*, i^*, h^*, p^*, a^*, \}$$

are given below

$$s^* = \frac{\delta i^* - (1-\varepsilon)\theta(i^* + p^* + h^* + a^*)[(1 - da^*)i^*] - \pi}{\pi - (1-\varepsilon)\theta(i^* + p^* + h^* + a^*) - da^*} - - - - - - - - - - - 10a$$

$$i^* = \frac{[(s^* + p^* + h^* + a^*)][(1-\varepsilon)\theta(p^* + h^* + a^*)] - da^*}{\left((s^* + p^* + a^* + h^*)(1-\varepsilon)\theta\right)} - - - - - - - - 10b$$

$$a^* = \frac{(h^* + p^*)\left(\pi + (1-\varepsilon)\theta(i^* + p^* + h^*)\right)}{(h^* + p^*)\left((1-\varepsilon)\theta - d\right) - (1 + \mu a)d + (1-\varepsilon)\left((i^* + p^* + h^*)\theta + 1\right)} - - - - - 10c \quad - - - - 10$$

$$h^* = \frac{\varepsilon_1 \delta i^* - ka^*}{1 - \left(\pi + (1-\varepsilon)\theta(i^* + p^* + h^* + a^*)\right)} - - - - - - - - - - - - 10d$$

$$p^* = \frac{vh^* - (\varepsilon_2 + \varepsilon_1)\delta i^* - h^*(\pi + (1-\varepsilon)\theta(i^* + p^* + a^*) - da^*)}{h(\pi - (1-\varepsilon)\theta) - (\gamma_2 + m) - (1-\varepsilon)\theta(i^* + p^* + h^* + a^*)} - - - - - - 10e$$

The fractional transformation of equation (5) in Caputo sense is

$$S(K+1) = \left[\frac{\Gamma\left(1+\frac{K}{\beta_1}\right)}{\alpha_1+1+\frac{K}{\beta_1}}\right][\pi - \frac{(\beta_{xx}+\beta_{xy}+\beta_{yx}+\beta_{yy}+\beta_{ax}+\beta_{xa}+\beta_{ay}+\beta_d+\beta_b+\beta_{jm}+\beta_f)}{N}[C_1\sum_{I=0}^{k} S(K)I(K-I) + C_2\sum_{I=0}^{k} S(K)P(K-I)$$

$$+ C_3\sum_{I=0}^{k} S(K)H(K-I) + C_4\sum_{I=0}^{k} S(K)A(K-I)] - S(K)\left(\pi + (1-\varepsilon)\theta\big(I(K)+P(K)+H(K)+A(K)\big) - dA(K)\right)] - - - - - - - 11a$$

$$I(K+1) = \left[\frac{\Gamma\left(1+\frac{K}{\beta_2}\right)}{\alpha_2+1+\frac{K}{\beta_2}}\right][\frac{(\beta_{xx}+\beta_{xy}+\beta_{yx}+\beta_{yy}+\beta_{ax}+\beta_{xa}+\beta_{ay}+\beta_d+\beta_b+\beta_{jm}+\beta_f)}{N}[C_1\sum_{I=0}^{k} S(K)I(K-I) + C_2\sum_{I=0}^{k} S(K)P(K-I)$$

$$+ C_3\sum_{I=0}^{k} S(K)H(K-I) + C_4\sum_{I=0}^{k} S(K)A(K-I)] + (1-\varepsilon)\theta(I(K)+P(K)+A(K)+H(K)) - \delta I(K)$$

$$- I\left(\pi + (1-\varepsilon)\theta\big(I(K)+P(K)+A(K)+H(K)\big) - dA(K)\right)] - - - - - - - - - - - - - - - - - - - - - - - - - - - - 11b$$

$$P(K+1) = \left[\frac{\Gamma\left(1+\frac{K}{\beta_3}\right)}{\alpha_3+1+\frac{K}{\beta_3}}\right][\varepsilon_2\delta I(K) - (1-m)\gamma_2 P(K) - m\gamma_1 P(K) - P(K)(\pi + (1-\varepsilon)\theta(I(K)+P(K)+A(K)+H(K)) - dA(K))] - - - - - 11c$$

$$A(K+1) = \left[\frac{\Gamma\left(1+\frac{K}{\beta_4}\right)}{\alpha_4+1+\frac{K}{\beta_4}}\right][(1-\varepsilon_1-\varepsilon_2)\delta I(K) + vH(K) + (1-m)\gamma_2 P(K) - kA(K) - dA(K) - A(K)(\pi$$
$$+(1-\varepsilon)\theta(I(K)+P(K)+A(K)+H(K)) - dA(K))] - - - - - - - 11d$$

$$H(K+1) = \left[\frac{\Gamma\left(1+\frac{K}{\beta_5}\right)}{\alpha_5+1+\frac{K}{\beta_5}}\right][\varepsilon_1\delta I(K) + m\gamma_1 P(K) + kA(K) - vH(K) - H(K)\left(\pi + (1-\varepsilon)\theta\big(I(K)+P(K)+A(K)+H(K)\big) - dA(K)\right)] - - - - 11e$$

(11)

$\beta_i, i = 1,2,3,4,5$ are the unknown values of the fractions $\alpha_i \quad i = 1,2,3,4,5$.

We have got

$S(K) = 0$ For K = 0,1,2,3,-------------$\alpha_1\beta_1 - 1$ $- - - - - - 12a$

$I(K) = 0$ For K = 0,1,2,3,-------------$\alpha_2\beta_2 - 1$ $- - - - - - 12b$

$P(K) = 0$ For K = 0,1,2,3,-------------$\alpha_3\beta_3 - 1$ $- - - - - - 12c$

$A(K) = 0$ For K = 0,1,2,3,-------------$\alpha_4\beta_4 - 1$ $- - - - - - 12d$

$H(K) = 0$ For K = 0,1,2,3,-------------$\alpha_5\beta_5 - 1$ $- - - - - - 12e$

12

POLYMATH package was used utilized on the data in table 2 and 3 and on equation (11) to obtain results.

## 4. ANALYSIS AND RESULTS

Table 2 and 3 shows values of parameters used in this paper

Table 2: Numerical data and parameter of people living with HIV/AIDS in Nigeria in 2015, Source: Avert [4]

| $Parameter$ | $Population$ | $\frac{Proportion}{N}$ $(Estimates\ in\ 4\ s.f)$ |
|---|---|---|
| $A$ | $220,000$ | 0.0129 |
| $I$ | $3.2\ million$ | 0.1031 |
| $d$ | 198 | 0.0039 |
| $P$ | $210,0000$ | 0.0258 |
| $T$ | $21\%\ of\ Infected\ people$ | 0.0790 |

Table 3: Numerical data for parameter , source: CATIE [6]

| Activities | Parameter | Data |
|---|---|---|
| Receptive vaginal sex | $\beta_{yx}$ | 0.6 |
| Insertive vaginal sex, male-to-female | $\beta_{xy}$ | 0.30 |
| Insertive anal | $\beta_{ax}$ | 0.27 |
| Receptive anal | $\beta_{xa}$ | 0.4 |
| Receptive fellatio | $\beta_f$ | 0.04 |
| Injecting drug use | $\beta_j$ | 2.4 |
| Needle-stick injury, no other risk factors | $\beta_m$ | 0.013 |
| Blood transfusion with contaminated blood | $\beta_d$ | 0.9 |
| Anal Sex (Man-Woman) | $\beta_{ay}$ | 0.6 |
| Anal Sex (Woman-Man) | $\beta_{ya}$ | 0.8 |
| Receptive Fellatio | $\beta_f$ | Negligible |

### 4.1. RESULTS OBTAINED VIA DIFFERENTIAL FRACTIONAL TRANSFORMATION

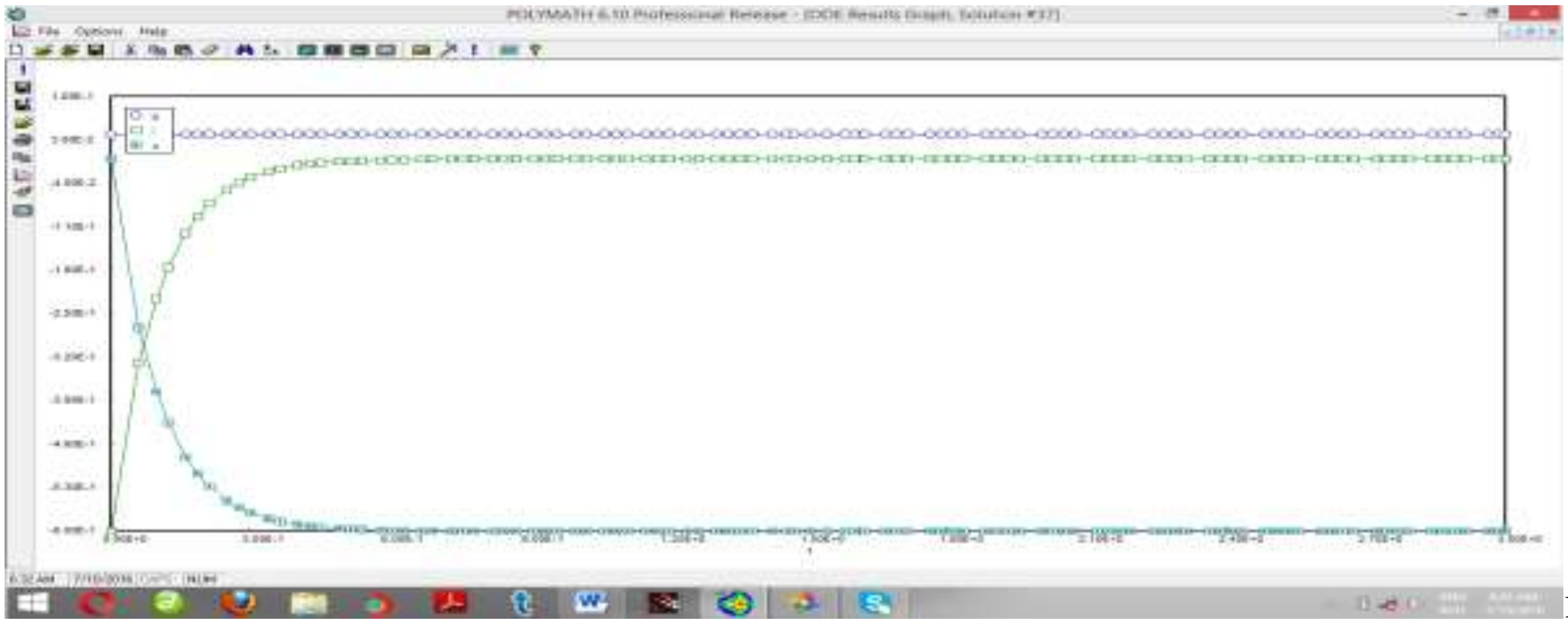

Figure 2: Result obtained when both treatment and adherence is not applied in Nigeria

Figure 2, shows the result obtained when both adherence and treatment is not applied

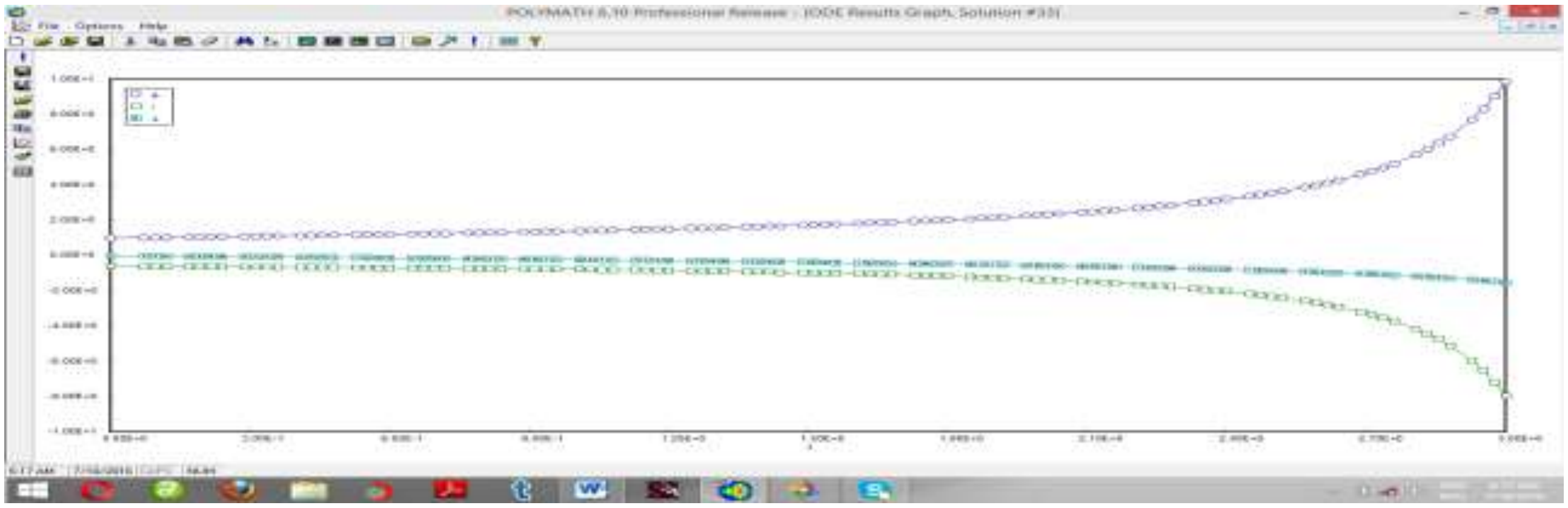

Figure 3: Result obtained when adherence is been applied in Nigeria

Figure 3, showing result showing result when adherence is applied but no treatment is applied. Susceptible population will raise, HIV infection will reduce and AIDS will be maintained at a constant level of control.

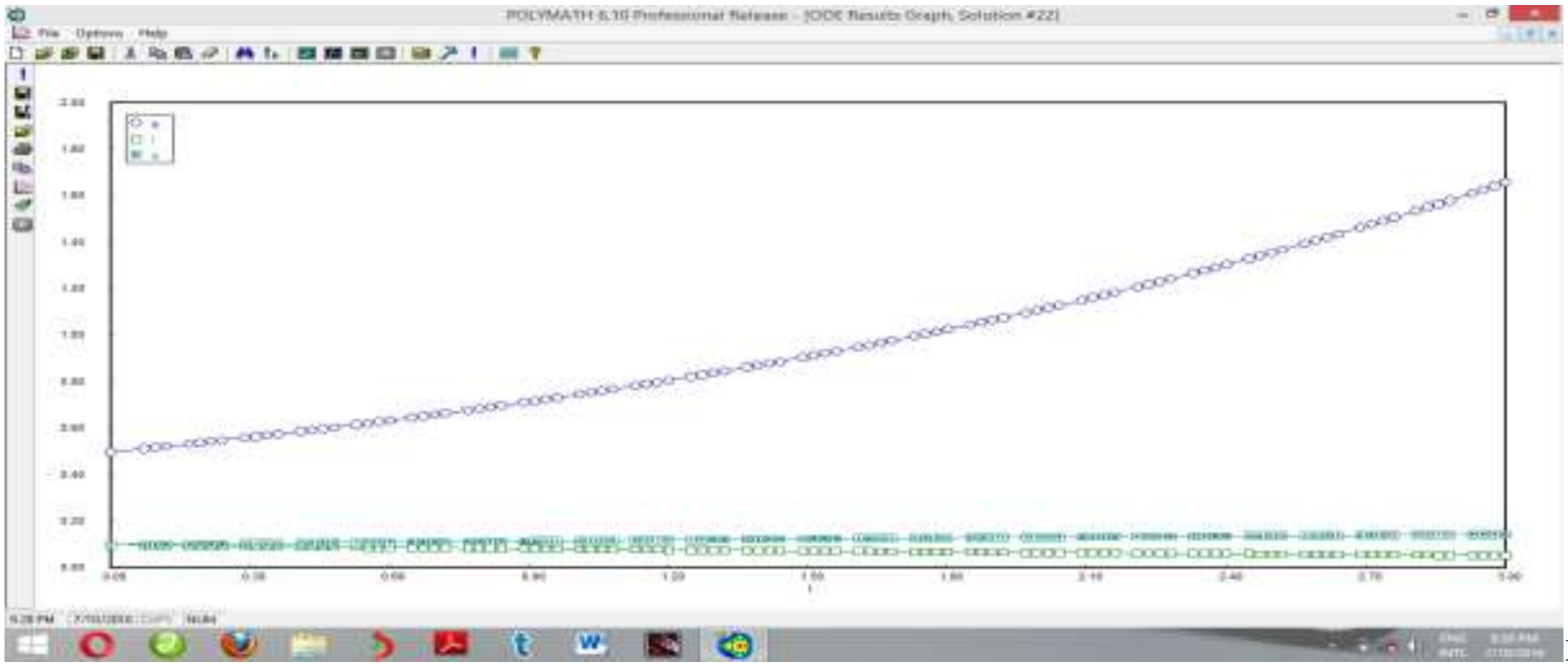
Figure 4: Result obtained when both treatment and adherence is applied in Nigeria

Figure 4, is the result obtained when both adherence and treatment is applied in Nigeria, susceptible population will continuously gradually increasing while HIV infection and AIDS infected population will be at the minimal growth.

### 4.2. MODEL VALIDATION VIA FRACTIONAL DIFFERENTIAL TRANSFORMATION

The steady states (equilibrium) values obtained by calculation are

$$\beta_1 = 0.3529, \quad \beta_2 = 0.2676, \quad \beta_3 = -2.5552, \quad \beta_4 = -0.6717 \quad \beta_5 = 0$$

Plugging these values into the solution equation of Runge-Kutta we obtain the final linear regression model.

$$N = a_1 S + a_2 I + a_3 P + a_4 A + a_5 T ------------ 4.1$$

Where the corresponding values of the $a_i$ i=1,2….,5 are as follows

Table 4: Final model parameter via differential transformation method

| Variable | Value | 95% confidence |
|---|---|---|
| a1 | 0.8295532 | 0.075262 |
| a2 | -0. 3062111 | 0.0321329 |
| a3 | 0.2013062 | 0. 1329032 |
| a4 | -0.4383062 | 0. 1322903 |
| a5 | 0.1253062 | 0. 2913032 |

### 4.3 FORECAST USING THE FINAL MODEL PARAMETER

Table 5: gives the forecast when differential transformation method is applied.

Table 5: Forecast using parameters in table 4

| $S$ | $I$ | $P$ | $A$ | $T$ | $N$ | $Forcast\ (Y_t)$ | $Residual\ (R)$ | $R^2$ |
|---|---|---|---|---|---|---|---|---|
| 0.5 | 0.5 | 0.5 | 0.5 | 0.5 | 0 | 0.000570503 | 0.000570503 | 3.25474E-07 |
| 0.6356676 | 0.4478335 | 0.324847 | 4.055602 | 45815 | 0.22047 | 0.220281262 | -0.000188738 | 3.5622E-08 |
| 0.6514715 | 0.4417586 | 0.3047344 | 5.312774 | 3.408962 | 0.2478131 | 0.247676009 | -0.000137091 | 1.87939E-08 |
| 0.6592882 | 0.4387543 | 0.2948093 | 6.082231 | 3.558873 | 0.2614776 | 0.261368535 | -0.000109065 | 1.18952E-08 |
| 0.6747547 | 0.4328104 | 0.2752159 | 7.974786 | 255443 | 0.2887951 | 0.28874969 | -4.54097E-05 | 2.06204E-09 |
| 0.6824057 | 0.4298706 | 0.2655453 | 9.13312 | 4.002129 | 0.3024491 | 0.302435335 | -1.3765E-05 | 1.89476E-10 |
| 0.697546 | 0.4240541 | 0.2464511 | 11.98213 | 4.292362 | 0.3297496 | 0.329798288 | 4.86882E-05 | 2.37054E-09 |
| 0.7050363 | 0.4211772 | 0.2370257 | 112585 | 4.435932 | 0.3433967 | 0.343475348 | 7.86481E-05 | 6.18553E-09 |
| 0.7198589 | 0.4154856 | 0.2184139 | 18.01463 | 4.720027 | 0.3706859 | 0.370818101 | 0.000132201 | 1.7477E-08 |
| 0.7271921 | 0.4126709 | 0.2092258 | 20.63954 | 4.860572 | 0.3843285 | 0.384482731 | 0.000154231 | 2.37873E-08 |
| 0.7417035 | 0.4071032 | 0.1910813 | 27.09565 | 5.138701 | 0.4116103 | 0.411800824 | 0.000190524 | 3.62992E-08 |
| 0.7488821 | 0.4043504 | 0.1821234 | 31.04704 | 5.276302 | 0.4252499 | 0.425452511 | 0.000202611 | 4.10513E-08 |
| 0.7630863 | 0.3989073 | 0.1644326 | 40.76568 | 5.548616 | 0.4525268 | 0.452741161 | 0.000214361 | 4.59507E-08 |
| 0.770112 | 0.3962173 | 0.1556984 | 46.71386 | 5.683344 | 0.4661644 | 0.466378278 | 0.000213878 | 4.57439E-08 |
| 0.7840104 | 0.3909014 | 0.138449 | 61.34368 | 5.949983 | 0.4934381 | 0.493639258 | 0.000201158 | 4.04647E-08 |
| 0.7908831 | 0.3882762 | 0.1299324 | 70.29768 | 6.081907 | 0.5070743 | 0.507262472 | 0.000188172 | 3.54087E-08 |
| 0.5267982 | 0.4896942 | 0.4650543 | 0.7275218 | 1.015231 | 0.041662 | 0.041836181 | 0.000174181 | 3.0339E-08 |
| 0.8375289 | 0.370563 | 0.0722779 | 182.4943 | 6.979974 | 0.6025209 | 0.602529102 | 8.20219E-06 | 6.72759E-11 |
| 0.8503638 | 0.3657437 | 0.0564179 | 239.6877 | 7.228597 | 0.6297898 | 0.629732374 | -5.74262E-05 | 3.29777E-09 |
| 0.8566929 | 0.3633816 | 0.0485871 | 274.6922 | 7.351612 | 0.6434242 | 0.643334134 | -9.00658E-05 | 8.11185E-09 |
| 0.8691641 | 0.3587635 | 0.0331215 | 360.787 | 7.595081 | 0.6706926 | 0.670546027 | -0.000146573 | 2.14835E-08 |
| 0.875301 | 0.3565132 | 0.0254856 | 413.4802 | 7.715547 | 0.6843267 | 0.684156648 | -0.000170052 | 2.89178E-08 |
| 0.8873609 | 0.3521475 | 0.010405 | 543.0813 | 7.953972 | 0.7115948 | 0.711391976 | -0.000202824 | 4.11377E-08 |
| 0.8932755 | 0.3500407 | 0.0029593 | 622.4018 | 8.071943 | 0.7252288 | 0.725018686 | -0.000210114 | 4.41481E-08 |
| 0.5951213 | 0.4634218 | 0.3767292 | 2.073674 | 2.327694 | 0.1519893 | 0.151756804 | -0.000232496 | 5.40543E-08 |
| 0.6033533 | 0.4602567 | 0.3661631 | 2.369895 | 2.485719 | 0.1657037 | 0.165467099 | -0.000236601 | 5.59799E-08 |
| 0.5529658 | 0.4796314 | 0.4310941 | 1.071582 | 1.518095 | 0.0831998 | 0.083127457 | -7.23427E-05 | 5.23346E-09 |
| 0.5700344 | 0.473068 | 0.4090303 | 1.392986 | 1.845974 | 0.1107675 | 0.110597703 | -0.000169797 | 2.8831E-08 |
| 0.5443214 | 0.4829556 | 0.4422948 | 0.9409326 | 1.352002 | 0.0693821 | 0.069376796 | -5.30363E-06 | 2.81285E-11 |
| 0.6196308 | 0.4539985 | 0.3453191 | 3.098514 | 2.798121 | 0.1931027 | 0.192879625 | -0.000223075 | 4.97623E-08 |
| 0.6276788 | 0.4509046 | 0.3350374 | 3.54448 | 2.952546 | 0.2067898 | 0.206581368 | -0.000208432 | 4.3444E-08 |
| 0.5784635 | 0.4698268 | 0.3981603 | 1.589726 | 2.007855 | 0.1245231 | 0.124323091 | -0.000200009 | 4.00036E-08 |
| 0.7977044 | 0.3856733 | 0.1214872 | 80.55958 | 6.212911 | 0.5207103 | 0.520882778 | 0.000172478 | 2.97486E-08 |
| 0.8111917 | 0.3805369 | 0.1048079 | 105.7992 | 6.472183 | 0.5479813 | 0.548109013 | 0.000127713 | 1.63105E-08 |
| 0.8178568 | 0.3780047 | 0.0965727 | 121.2467 | 6.600465 | 0.5616164 | 0.561717247 | 0.000100847 | 1.01701E-08 |
| 0.8310263 | 0.3730166 | 0.0803083 | 159.2407 | 6.854354 | 0.5888862 | 0.588927066 | 4.08663E-05 | 1.67006E-09 |

### 4.4. ANOVA DATA TABLE

The ANOVA data value is shown below

Table 6: ANOVA data value

| | |
|---|---|
| Residual Square ($R^2$) | 0.7517414 |
| Adjusted ($R^2$) | 0.7492081 |
| Root Mean Square Deviation | 0.0166519 |
| Variance | 0.0282943 |

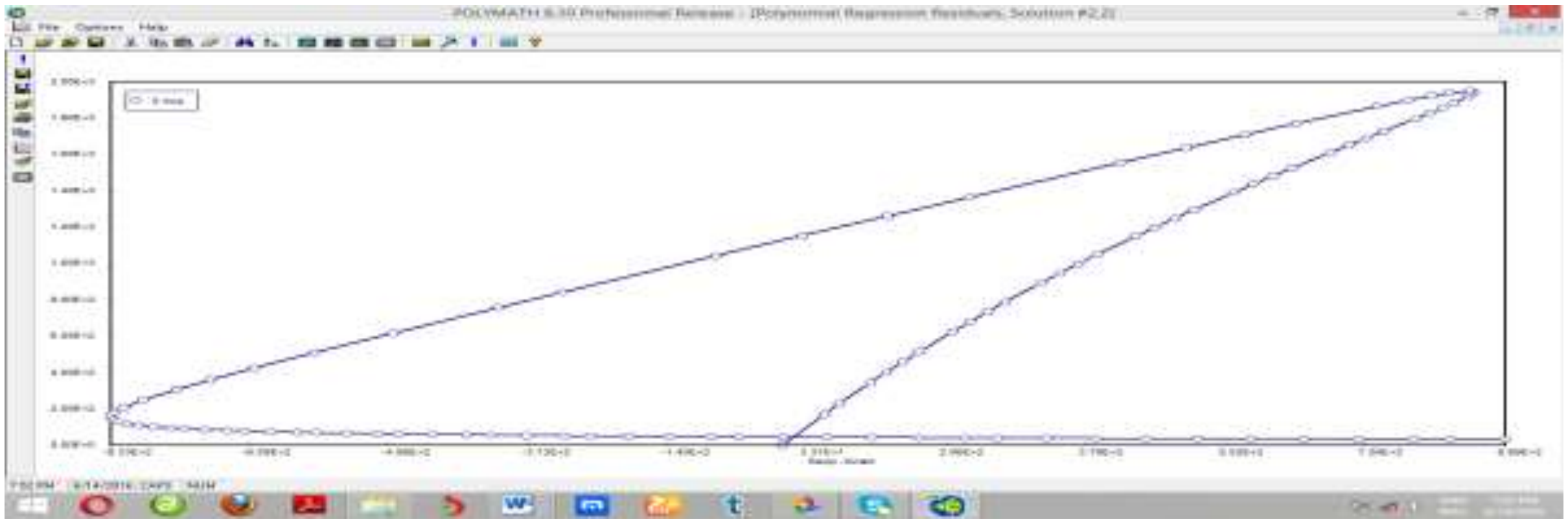

Figure 4.11: Forecast error of the model

Figure 4.11 showing result of the forecast error

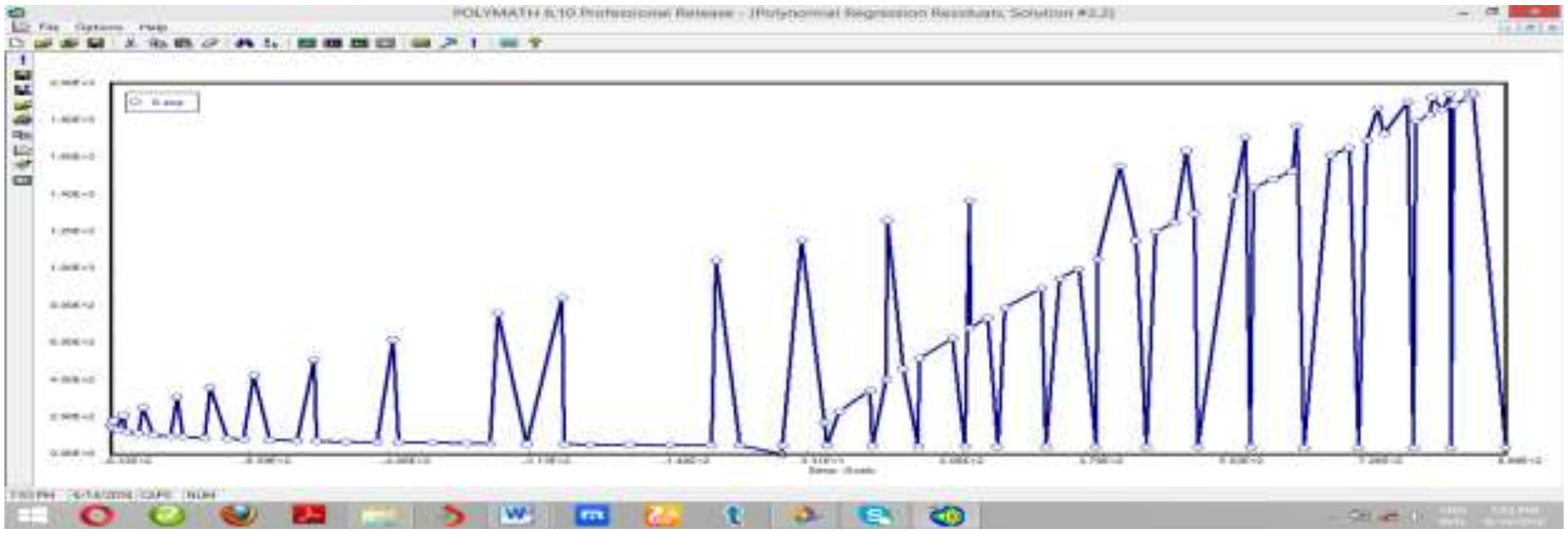

Figure 4.12: graph of actual data plot

Figure 4.12 is graph showing result of the actual data plot

## 5. SUMMARY

In this work, a nonlinear model was proposed and analyzed to study the dynamic of HIV/AIDS transmission in a time-varying population size using a network transitional diagram in which resulted to the analytical model. The analytical model framework was tested to see if it satisfies epidemiological model (positivity solution). Numerical solution of the model was conducted via Fractional Differential Transformation by Caputo. A graphical representation of the result was given. The model was tested to determine its goodness of fit via linear regression analysis.

## 5.1 DISCUSSION

From the result obtained HIV/AIDS can be controlled efficiently if both treatment and adherence is applied in Nigeria. Applying only treatment without adherence will control the spread of the virus however the desired goal of the government and Non-governmental organization (NGO) will not be successfully achieved and vice versa. In situation where both of this is not applied the virus will rapidly grow.

The $R^2$ obtained from Fractional Differential transformation (Caputo) method was larger (i.e, 0.7517414) which is approximately 75%. Similarly the plots of the observed regression versus fitted response of the two methods also show that the model captures all the variability of the responds data around its mean.

## 5.2 CONCLUSSION

A modified model from the previous work was formulated and tested. The proposed model was used for forecast. The forecast was analyzed, giving accuracy of 75%. This would provide a useful managemanet and control tool. The government should provide more incentives in terms of drug subsidy, information, awareness, education and periodic check on the citizens.

## 5.3 CONTRIBUTIONS

1) This study expands on some previous studies of mathematical models of HIV/AIDS
2) This research work informs how the use of ARV treatment can control and possibly eradicate HIV/AIDS.
3) Provides a generalized framework for analytical and further research activities towards eradicating HIV/AIDS
4) Provides more general case for handling many cases of HIV/AIDS diagnosis and treatments.